\documentclass[prl,twocolumn]{revtex4}
\usepackage{epsfig}
\usepackage{amsmath}
\usepackage{amsfonts}

\begin{document}

\title{Coulomb drag as a measure of trigonal warping in doped graphene}

\author{B.N. Narozhny}

\affiliation{The Abdus Salam ICTP, Strada Costiera 11, Trieste, I-34100, Italy}

\date{\today}

\begin{abstract}
  I suggest to use the effect of Coulomb drag between two closely
  positioned graphite monolayers (graphene sheets) for experimental
  measurement of the strength of weak non-linearities of the spectrum
  in graphene. I consider trigonal warping as a representative
  mechanism responsible for the drag effect. Since graphene is
  relatively defect-free, I evaluate the drag conductivity in the
  ballistic regime and find that it is proportional to the fourth
  power of the warping strength.
\end{abstract}

\maketitle

The first experimental measurement of conducting properties of
graphene \cite{gr1} (an atomically thin crystalline monolayer of
graphite) was followed by developing graphene-based transistors
\cite{gr2,gr3}, where high concentrations of charge carriers can be
induced by applying gate voltages. These discoveries have brought a
lot of attention to the field, which is well-studied theoretically.
Indeed, a two-dimensional, hexagonal lattice of carbon atoms is a
usual starting point for most calculations on bulk graphite, 
carbon nanotubes \cite{dre}, or fullerenes \cite{re3}. In
fact it's been almost 60 years since the band structure of graphene
has been first studied \cite{th1}. 

The continuous theoretical interest in graphene is due to the
Dirac-type dispersion relation \cite{spe,sem} leading to a number of
peculiar properties -- from the Berry phase of electronic
wavefunctions \cite{ber} to an anomalous Quantum Hall Effect
\cite{gr3}. The chiral nature of charge carriers in graphene is a
consequence of its crystal structure. The honeycomb lattice contains
two equivalent sublattices. Nearest-neighbor hopping between $A$ and
$B$ sites results in formation of two energy bands which intersect
near the corners of the hexagonal Brillouin zone \cite{dre}. Two
inequivalent corners $K_\pm$ \cite{kpm} define a valley index,
specifying excitations around the Fermi energy (playing a role of
a pseudospin). Close to the crossing points the spectrum is conical
with the ``light-speed'' $v\approx 10^6$ m/s \cite{gr2}.

{
\begin{figure}[ht]
\epsfxsize=6 cm
\centerline{\epsfbox{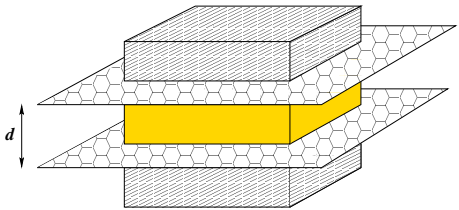}}
\caption{A Coulomb drag sample: two graphene sheets are
  separated by an insulating layer of the thickness $d$. The two gates
  at the top and bottom of the device can be used to independently
  control carrier concentrations in the two sheets.}
\label{ex}
\end{figure}
}

The true microscopic Hamiltonian in graphene contains several small
corrections to the Dirac spectrum. For example, using a second order
${\bf k}\cdot{\bf p}$ equation \cite{aan} one can derive a quadratic
term in the effective low-energy Hamiltonian of graphene that violates
the isotropy of the Dirac spectrum and causes trigonal warping
\cite{ber}. When gate voltage is applied, resulting in
non-zero Fermi energy, the electronic Fermi line
deviates from a perfect circle: the ${\bf p} \rightarrow -{\bf p}$
symmetry of the Fermi surface is broken within each valley. Although
weak, such distortions of the Fermi surface lead to observable
effects: trigonal warping suppresses antilocalization \cite{fwl},
which one would otherwise expect due to the absence of backscattering
\cite{ber,hln}.

Trigonal warping also breaks another, more subtle symmetry in the
problem -- the symmetry between excitations above and below the finite
Fermi energy $E_F = \hbar v\sqrt{\pi n_e}$ ($n_e$ being the electron
density). Such asymmetry can be detected in transport measurements. A
particular ``tell-tale'' experiment that crucially depends on its
presence is Coulomb drag \cite{dr1,dr2,dt1,dt2,dt3,kam,dt4}. Coulomb
drag measurements are performed on two closely positioned (but
electrically isolated) layers. A current $I_a$ driven through one of
the layers (the ``active'' layer) induces a voltage drop $V_p=\rho_D I_a$
in the other (``passive'') layer. The voltage appears due to
inter-layer electron-electron interaction that creates a frictional
force ``dragging'' electrons in the passive layer. As a purely
interaction effect, Coulomb drag has become a sensitive tool for
experimental studies of electron-electron interaction in many problems
of contemporary condensed matter physics. It has been used in search
for Bose condensation of interlayer excitons \cite{dr3}, a
metal-insulator transition in two-dimensional layers \cite{dr4}, and
Wigner cristallization in quantum wires \cite{dr5}.

Other effects contributing to non-linearity of the spectrum in
graphene will also result in non-zero contribution to the Coulomb
drag. These include: the quadratic correction to quasiparticle
spectrum due to next-neighbor hopping \cite{dre}; Coulomb scatterers
resulting in energy-dependent scattering time $\tau_c\propto|E_F|$
\cite{mac}; logarithmic corrections to quasiparticle spectrum
\cite{abr}; and interference corrections to scattering time
\cite{anr}. In this Letter I consider trigonal warping as a
representative mechanism of the Coulomb drag in graphene, leaving the
discussion of the role of other mechanisms for a subsequent
publication \cite{me1}. I argue that this effect can be distinguished
from other drag mechanisms by its dependence on inter-layer separation
$d$ and Fermi momenta (or gate voltages).

I envision a following set-up (see Fig.~\ref{ex}). Two graphene sheets
are positioned parallel to each other and are separated by an
insulating material about $50$ nm thick (e.g. using the technique
recently developed in Ref.~\onlinecite{gor}). Two gates (at the top
and bottom of the device) can be used to independently control carrier
concentrations in the two layers. I assume that sufficient gate
voltage is applied, so that Fermi energies in each layer are positive
$E_F^{(a,p)}>0$ and represent the largest energy scale in the problem
(indices $a$ and $p$ denote the active and passive layers). Since
graphene is relatively defect-free, the elastic scattering rate
$\tau^{-1}$ is assumed to be much smaller than temperature. At the
same time, the mean-free path $\ell$ is the longest length scale. The
assumed hierarchy of energy and length scales (here $\lambda_F$ is the
Fermi wavelength)
\begin{equation}
\hbar\tau^{-1}\ll T\ll E_F^{(a,p)}, \quad
\ell^{(a,p)} \gg d \gg \lambda_F^{(a,p)},
\label{ls}
\end{equation}
\noindent
ensures that the the device is in the ballistic regime \cite{ffo} (for
Coulomb drag in the ballistic regime see Ref.~\onlinecite{kam}).

At lower temperatures ($T<\hbar\tau^{-1}$) electron motion becomes
diffusive. In that case, scattering off atomically sharp disorder
becomes important for it breaks the pseudospin symmetry and
drastically affects two-particle correlation functions \cite{fwl}. An
analysis of the Coulomb drag in the diffusive regime will be
considered elsewhere \cite{me1}.

In the weak coupling regime \cite{dt3,kam,dt4} the drag coefficient
is proportional to the drag conductivity
$\rho_D\approx\sigma_D/(\sigma_a\sigma_p)$ ($\sigma_{(a,p)}$ being the
Drude conductances of the two layers). The latter is typically
calculated using the expresssion
\begin{equation}
\label{sd}
\sigma^{\beta\beta'}_D=\frac{1}{8 TS}\sum\limits_{\bf q}
\int\frac{d\omega}{2\pi}
\left|{\cal D}_{ap}\right|^2
\frac{
\Gamma^\beta_a(\omega, {\bf q})
\Gamma^{\beta'}_p(\omega, {\bf q})
}
{
\sinh^2\frac{\omega}{2T}
}
,
\end{equation}
\noindent
where $S$ is the sample area, ${\cal D}_{ap}$ is the screened
inter-layer interaction, and ${\bf \Gamma}$ is the non-linear
susceptibility (or rectification function) that relates a scalar
potential $V( \omega, {\bf q})$ to the {\it dc} current it creates in
quadratic response ${\bf j}_{dc} = {\bf \Gamma(\omega, {\bf q})} |V(
\omega, {\bf q})|^2$.  Below I re-derive Eq.~(\ref{sd}) for graphene
in the ballistic regime under the above assumptions and show, that for
Dirac particles the drag vanishes.

When trigonal warping is taken into account, I find that the drag
conductivity is proportional to the fourth power of the parameter
$W$ that describes the strength of the warping correction to 
the Dirac spectrum:
\begin{equation}
\label{res}
\sigma_D = \frac{e^2}{\hbar}
\frac{5\zeta[5]}{32}
\frac{\hbar^2}{(\kappa_a d)(\kappa_p d)}
\frac{T^2W^4}{v^6d^2}
\frac{\ell_a\ell_p}{d^2}.
\end{equation}
\noindent
Here $\kappa_{(a,p)}=e^2 k_F^{(a,p)}/v$ are the Thomas-Fermi momenta.
Eq.~(\ref{res}) is the main result of this communication.

The low-energy single-particle Hamiltonian \cite{spe,ber,fwl}
in graphene can be written in the basis
of $4$-component Bloch functions $\Phi =(\phi_{A, K_+},
\phi_{B, K_+}, \phi_{B, K_-}, \phi_{A, K_-})$ (I employ notations
introduced in Ref.~\onlinecite{fwl}; $\phi_{A, K_+}$ is
the electronic amplitude on sublattice $A$ and valley $K_+$) as
\begin{equation}
\label{h0}
\widehat{\cal H}_0 = v\vec\Sigma {\bf p}
-W\Big[ \hat\sigma_x \left( p_x^2-p_y^2\right)
-2\hat\sigma_y p_x p_y \Big],
\end{equation}
\noindent
with the weak quadratic term causing trigonal warping. Here Pauli
matrices $\hat\sigma_i$ act in the sublattice space $(A, B)$. The
``isospin'' $\vec\Sigma$ is defined as direct products of Pauli
matrices ${\bf \hat\sigma}$ (acting in the sublattice space) and ${\bf
  \Pi}$ (acting in the valley space $K_\pm$):
$\Sigma_{x(y)}=\Pi_z\otimes\hat\sigma_{x(y)}$.

In the basis of plane waves $\widehat {\cal H}_0$ is a $4\times 4$
matrix that can be diagonalized by a unitary transformation 
$\hat R^{-1}\widehat{\cal H}_0\hat R=
{\rm diag}[E_{\alpha,\xi}({\bf p})]$.
The resulting eigenvalues are
\begin{equation}
\label{evwf}
E_{\alpha,\xi} = \alpha vp s_\xi; \; s_\xi=
\sqrt{1-2\xi pWv^{-1}\cos3\varphi_{\bf p} +p^2W^2v^{-2}},
\end{equation}
\noindent
where $\xi = \pm 1$ denotes the two valleys, $\alpha = \pm 1$ is the
chirality and distinguishes between the conductance ($\alpha = 1$) and
valence ($\alpha = -1$)bands, and $\varphi_{\bf p}$ is an angle between
the momentum ${\bf p}$ and the $x$-axis ($\tan\varphi_{\bf p} = p_y/p_x$).

The electron field operator can be written in the basis of the
eigenstates as (hereafter I use the units with $\hbar=1$)
\begin{widetext}
\begin{eqnarray}
\label{psi}
\widehat \Psi ({\bf r}) =\sum\limits_{{\bf p}, \alpha, \xi}
\psi_{{\bf p}, \alpha, \xi}({\bf r}) \widehat a_{{\bf p}, \alpha, \xi}=
\frac{1}{\sqrt{2}}\sum\limits_{{\bf p}, \alpha}e^{i{\bf p r}}
\begin{pmatrix}
e_{\alpha, \xi=1} \cr 1 \cr 0 \cr 0 \cr
\end{pmatrix}
\widehat a_{{\bf k}, \alpha, \xi=1} +
\frac{1}{\sqrt{2}}\sum\limits_{{\bf p}, \alpha}e^{i{\bf p r}}
\begin{pmatrix}
0 \cr 0 \cr e_{\alpha, \xi=-1} \cr 1 \cr 
\end{pmatrix}
\widehat a_{{\bf p}, \alpha, \xi=-1},
\end{eqnarray}
\end{widetext}
\noindent
where
$e_{\alpha, \xi}= \alpha s_\xi e^{2i\varphi_p}/[e^{3i\varphi_p}-(\xi pW/v)]$.
Then the form of the electron density operator 
 \begin{equation}
\label{rho}
\widehat\rho({\bf r}) = \sum\limits_{{\bf p},{\bf p}',\alpha, \alpha',\xi}
 e^{-i({\bf p}-{\bf p}'){\bf r}} \widehat a^\dagger_{{\bf p}, \alpha, \xi}
\widehat a_{{\bf p}', \alpha', \xi}\lambda^{\alpha,\alpha'}_{{\bf p},{\bf p}'},
\end{equation}
differs from the usual one by the presence of vertices 
\begin{equation}
\label{l}
\lambda^{\alpha,\alpha'}_{{\bf p},{\bf p}'}=\left(1+e_{\alpha, \xi} e^*_{\alpha', \xi}\right)/2.
\end{equation}
\noindent
The vertices $\lambda^{\alpha,\alpha'}_{{\bf k},{\bf k}'}$ indicate the
asymmetry of quasi-particle scattering in graphene. In particular,
the supression of backscattering \cite{ber} follows from the fact 
that in the absence of trigonal warping 
$\lambda^{\alpha,\alpha}_{{\bf p},-{\bf p}}(W=0)=0$. Similarly,
$\lambda^{\alpha,-\alpha}_{{\bf p},{\bf p}}(W=0)=0$,
indicating that inter-band transitions at the same wave-vector
are also supressed.

The single-particle Green's function in the original basis of 
Bloch functions is a $4\times 4$ matrix \cite{fwl} 
It can be brought to a diagonal form by
the rotation $\hat R$. In any closed loop such operation would cause
the appearance of the vertex factors (\ref{l}). For example, the
polarization operator (the density-density response function) is
\begin{eqnarray}
\label{p0}
&&
{\cal P}_0(\omega, {\bf q}) = -2i\sum\limits_{\alpha, \alpha',\xi}
\int\frac{d^2k}{(2\pi)^2} \int\frac{d\epsilon}{2\pi}
\\
&&
\nonumber\\
&&
\quad\quad\quad\quad
\times\left|\lambda^{\alpha, \alpha'}_{{\bf k}, {\bf k}+{\bf q}}\right|^2
G_{\alpha,\xi}(\epsilon, {\bf k}) \;
G_{\alpha',\xi}(\epsilon + \omega, {\bf k}+{\bf q}),
\nonumber
\end{eqnarray}
\noindent
where the overall factor of $2$ follows from the spin degeneracy. Here
$G_\alpha(\epsilon, {\bf k})$ is the Green's function for the $\alpha$
band. For $E_F>0$ and $T\ll E_F$ one can use the standard
non-relativistic Green's functions \cite{chi}.

Since the vertex functions are frequency-independent, the usual
reasoning \cite{mbp} leads to the following expression for the
retarded polarization operator in the ballistic regime
($f[E_\alpha({\bf k})]$ is the Fermi function)
\begin{eqnarray}
\label{pr}
&&
{\cal P}^R_0(\omega, {\bf q}) = -2\sum\limits_{\alpha, \alpha', \xi}
\int\frac{d^2k}{(2\pi)^2}
\\
&&
\nonumber\\
&&
\quad\quad\quad
\times
\left|\lambda^{\alpha, \alpha'}_{{\bf k}, {\bf k}+{\bf q}}\right|^2
\frac{f[E_{\alpha',\xi}({\bf k}+{\bf q})] - f[E_{\alpha,\xi}({\bf k})]}
{\omega - E_{\alpha',\xi}({\bf k}+{\bf q}) + E_{\alpha,\xi}({\bf k}) + i\eta},
\nonumber
\end{eqnarray}
\noindent
where $\eta\rightarrow +0$. Note, that under assumptions that the
Fermi level is in the conduction band and $T\ll E_F$, the term with
$\alpha=\alpha'=-1$ (the valence band contribution) vanishes due to
the Fermi functions.

For the purposes of describing screened inter-layer interaction in the
Coulomb drag problem, I am only interested in momenta $q$ smaller than
inverse inter-layer distance. Under the assumption (\ref{ls}), 
$q < (1/d) \ll k_F$.
Thus, the inter-band contribution to Eq.~(\ref{pr}) is suppressed by a
factor of $q^2/k_F^2$ and the polarization operator is dominated by
the conduction band. Due to the Fermi functions in Eq.~(\ref{pr}) the
momentum integral is dominated by the region $k\sim k_F$. Then the
leading contribution to the static polarization operator needed in the
Coulomb drag problem in the ballistic regime \cite{kam} is
\begin{equation}
\label{prr}
{\cal P}^R_0(\omega=0) = 2k_F/\pi v = 4\nu.
\end{equation}
\noindent
Here $\nu$ is the density of states at the Fermi level (spin and
valley degeneracy is taken into account). Consequently, the screened
interlayer interaction is the same as in the case of the uniform
two-dimensional electron gas \cite{kam}:
\begin{equation}
\label{d}
{\cal D}_{ab} = \pi e^2 q/(\kappa_a \kappa_p \sinh qd).
\end{equation}

To derive the expression for the drag conductivity (\ref{sd}) one
starts with the general expression for electric current in the passive
layer in terms of the Keldysh Green's function. In graphene the
current vertex (in the original basis of Bloch functions) is ${\bf
  \widehat J} = 2ev\vec\Sigma$. Diagonalizing the Green's function by
a unitary transformation the current in the passive layer takes the
form
\begin{equation}
{\bf j} = - \frac{i}{2S} \sum\limits_{\alpha, \xi} \int d^dr
{\bf \hat j}_{\alpha, \xi}
G^K_{\alpha, \xi}; \;
{\bf \hat j}_{\alpha, \xi} = 
\left(\hat R^{-1}{\bf \widehat J}\hat R\right)_{\alpha,\alpha}^{\xi,\xi}.
\label{cdt}
\end{equation}
\noindent
In a system of free electrons in equilibrium the current (\ref{cdt})
is equal to zero. Perturbing the Green's function by a potential
$V(\omega)$ one finds the following expression for the current [more
precisely, the contribution relevant for the drag problem; here I use
a short-hand notation for the spatial coordinates in the argument of
Green's functions -- $G^R_{\alpha, \xi} (\epsilon; 31) = G^R_{\alpha,
  \xi} (\epsilon; {\bf r}_3,{\bf r}_1)$]
\begin{widetext}
\begin{equation}
{\bf j} = \frac{i}{V} \sum\limits_{\alpha, \xi}
\int d^dr_3 \int \frac{d\epsilon}{2\pi}
\frac{d\omega}{2\pi} \Big(
f[\epsilon] - f[\epsilon- \omega]\Big)V(\omega; 12) 
\left[\sum\limits_{\alpha'}
    {\rm Im} G^R_{\alpha', \xi}(\epsilon - \omega; 12) 
\left| \lambda^{\alpha, \alpha'} \right|^2
\right]
G^A_{\alpha, \xi}(\epsilon; 23){\bf \hat j}_{\alpha, \xi} 
G^R_{\alpha, \xi} (\epsilon; 31).
\label{j1}
\end{equation}
\end{widetext}

In the ballistic regime \cite{kam} Green's functions in Eq.~(\ref{j1})
are averaged over the disorder independently of each other. Then one
can perform the Fourier transform and use the approximation ($\tau$
is the elastic scattering rate):
\begin{subequations}
\label{gdel}
\begin{equation}
\label{grga}
G^R_{\alpha, \xi} (\epsilon; {\bf k}) 
G^A_{\alpha, \xi} (\epsilon; {\bf k}) \approx 
2\pi \tau \delta
\Big( \epsilon - E_{\alpha, \xi}({\bf k}) \Big),
\end{equation}
\begin{equation}
\label{imgr}
{\rm Im} G^R_{\alpha, \xi} (\epsilon; {\bf k}) 
\approx -\pi
\delta\Big( \epsilon - E_{\alpha, \xi}({\bf k}) \Big).
\end{equation}
\end{subequations}
\noindent
In this case 
\[
f[\epsilon] - f[\epsilon- \omega] \rightarrow
f[E_{\alpha, \xi}({\bf k})] 
- f[E_{\alpha, \xi}({\bf k}-{\bf q})].
\]
\noindent
It is then clear that if the Fermi level is in the conduction band
then the two Fermi functions for the valence bands cancel each
other. Inter-band processes are suppressed similarly to the case of
the polarization operator and will be neglected hereafter. Thus, in
the ballistic regime, only particles in the conduction band contribute
to the current in the passive layer, as one would intuitively
assume. The question of whether this statement remains true when
off-diagonal disorder is taken into account, i.e. in the diffusive
regime, will be discussed in a subsequent publication \cite{me1}. The
situation in the active layer is similar, in fact both layers are
described by the same non-linear susceptibility ${\bf \Gamma}$.
Therefore, the general expression for the drag conductivity (\ref{sd})
remains valid (again, with only particles in the conduction band
contributiong). What remains to be done inorder to obtain the result
(\ref{res}) is to evaluate the non-linear susceptibility.

Under my assumptions the non-linear susceptibility for the conduction 
band in the ballistic regime is
\begin{equation}
{\bf \Gamma}(\omega, {\bf q}) = 
\frac{\omega}{\pi}
\left[ {\vec \gamma}_{\bf q}(\omega)
- {\vec \gamma}_{-{\bf q}}(-\omega)
\right],
\label{g}
\end{equation}
\noindent
where the triangular vertex ${\vec \gamma}$ is given by
\begin{eqnarray}
\label{gv}
&&
{\vec \gamma}_{{\bf q}}(\omega) = \sum\limits_\xi
\int\frac{d^2k}{(2\pi)^2} 
\left| \lambda^{1, 1}_{{\bf k}, {\bf k}+{\bf q}} \right|^2
\\
&&
\nonumber\\
&&
\quad\quad
\times
\Big[{\rm Im}
G^R_{1,\xi}(\epsilon + \omega; {\bf k}+{\bf q})\Big] 
G^{R}_{1,\xi}\left(\epsilon; {\bf k}\right) 
{\bf \hat j}_1
G^{A}_{1,\xi}\left(\epsilon; {\bf k}\right).
\nonumber
\end{eqnarray}
\noindent
Here ${\bf \hat j}_1$ is the diagonal matrix
element of the curent vertex $\widehat{\bf J}$ rotated (\ref{cdt}) to the 
basis of the eigenfunctions: ${\bf \hat j}_1 = 2ev {\bf n}_{\bf k}$.
Note how the current vertex for the conduction band recovers its usual
momentum dependence!

Consider now the Coulomb drag for Dirac particles. Setting $W=0$ in
Eq.~(\ref{l}), one finds [$\theta=\angle({\bf k}, {\bf q})$]
\begin{eqnarray}
\label{i1}
&&
\gamma^\beta = -4ev_F \pi \nu \tau_0
\int\frac{d\theta}{2\pi} n^\beta_{\bf k}
\left(1-\frac{q^2}{4k^2}\sin^2\theta\right)
\\
&&
\nonumber\\
&&
\quad\quad\quad
\times
\delta\Big(\omega-E_1({\bf k}+{\bf q})+E_1({\bf k})\Big).
\nonumber
\end{eqnarray}
\noindent
The result is even under the simultaneous change of sign of $\omega$ 
and ${\bf q}$: 
${\vec \gamma}_{{\bf q}}(\omega) = {\vec \gamma}_{-{\bf q}}(-\omega)$.
Therefore ${\bf\Gamma}(\omega,{\bf q}) = 0$. There is no drag
effect in a system with linear spectrum.

When the deviation from linearity in Eq.~(\ref{h0}) is taken into
account, the contribution of the two valleys to Eq.~(\ref{i1}) is no
longer identical and a non-zero result appears only in the second
order in $W$:
\begin{equation}
\label{gres}
{\bf \Gamma} (\omega, {\bf q}) =
-4e\nu {\bf q}  \omega w^2\ell v^{-3} \sin\varphi_{\bf q}
\cos 3\varphi_{\bf q}
\theta(vq-\omega),
\end{equation}
\noindent
where the mean-free path is defined as $\ell = 2v\tau$. Using 
Eqs.~(\ref{gres}) and (\ref{d}) in Eq.~(\ref{sd}), I find the 
final result (\ref{res}).

To summarize, I have considered the Coulomb drag between two closely
positioned graphene sheets. For strictly linear Dirac-type
dispersion, I find that the drag vanishes, in agreement with the
traditional interpretation of the effect as a manifestation of
asymmetry between elementary excitations above and below the Fermi
level. As a representative mechanism of such asymmetry in graphene I
consider trigonal warping and find the drag coefficient proportional
to the fourth power of the strength $W$ of the warping term
\cite{ftn}. The obtained result should be distinguishable from the
drag due to other non-linear contributions \cite{dre,mac,abr,anr} to
graphene spectrum by its dependence on inter-layer separation and
Fermi momentum. In my opinion the Coulomb drag is an ideal tool for
experimental studies of spectrum non-linearities in graphene.

I am grateful to V.I Fal'ko for attracting my attention to this
problem, to I.L. Aleiner for helpful discussions and to MPI-PKS,
Dresden for hospitality during the workshop ``Dynamics and Relaxation
in Complex Quantum and Classical Systems and Nanostructures'' (2006).


\begin{thebibliography}{99}
\bibitem{kpm} The two inequivalent corners of the Brillouin zone
  $K_\pm$ are $(\pm\frac{4}{3}\pi a^{-1},0)$, where $a$ is the lattice
  constant.
\bibitem{ftn} I believe that the role of the symmetry between excitations
above and below $E_F$ was ovelooked in Ref.~\onlinecite{srp}.
\end{thebibliography}
\end{document}